\documentstyle[preprint,aps]{revtex}

\begin{document}
\draft
\title{
New Types of Off-Diagonal Long Range Order in Spin-Chains
}
\author{
J. C. Talstra$^{(1)}$,
 S. P. Strong$^{(2)}$
 and P. W. Anderson$^{(1)}$}

\address{$^{(1)}$Joseph Henry Laboratories,
Jadwin Hall,
P.O. Box 708, Princeton, NJ 08544-0708\\
$^{(2)}$NEC Research Institute,
4 Independence Way, Princeton, NJ 08540
}
\date{\today}
\maketitle

\begin{abstract}
We discuss new possibilities for Off-Diagonal Long Range Order
(ODLRO) in
spin chains involving operators which add or delete sites from the
chain.  For the Heisenberg and Inverse Square Exchange
models we give
strong numerical evidence for the hidden ODLRO
conjectured by Anderson \cite{pwa_conj}.  We
 find a similar ODLRO for the XY model
(or equivalently for free fermions in one spatial dimension) which
we can demonstrate rigorously, as well as numerically.
A connection to the singlet pair correlations in
one dimensional
models of interacting electrons is made and briefly discussed.

\end{abstract}
\pacs{}

In 1991 one of us (P.W.A) conjectured, based on RVB ideas for
the one-dimensional Hubbard model,
that there should be a non-zero overlap between the groundstate of
the one dimensional Heisenberg model on a chain of $N$ sites
and the state obtained by inserting a pair of nearest neighbor
spins in a singlet configuration into the ground state of the
$N-2$ site Heisenberg chains \cite{pwa_conj,Hald88}.
Both of these models were therefore
expected to have a hidden form of ODLRO for an operator which
not only involved sampling the state of the system (in this case
checking that a given pair of spins was in a singlet) but also
changing the Hilbert space of the system in an essential way
(adding two sites),
reminiscent of the Girvin-MacDonald-Read order parameter of the
Fractional Quantum Hall Effect (FQHE) \cite{FQHE}, despite the
fact that that the Heisenberg model has gapless excitations.
The conjecture that the overlap should be
finite has not been previously
investigated (but see \cite{PS} where
evidence for the resulting ODLRO
was found);
%
%
therefore
we numerically tested the original conjecture
that the overlap for the ISE and Heisenberg models
 between the $N$ site groundstate
and the $N-2$ site groundstate with a local singlet inserted
should be non-zero.

%
%
These
%
%
overlaps as a function of
$N$ are shown in Fig. \ref{fig:size}
together with  fits to the results of the form
$0.817 + 0.778 N^{-2}$ for the ISE model and
$0.820 + .740 N^{-2}$ for the Heisenberg model.
The results strongly suggest that both overlaps
remains finite in the limit as the system size
goes to infinity.  Note that since the
phases of the $N$ and $N-2$ site wavefunctions
may be chosen independently the phase of the overlaps is
meaningless and further, since the groundstate momenta
of the $N$ and $N-2$ site groundstate wavefunctions
differ by $\pi$, the overlap is multiplied by minus one
if the location of the singlet pair is shifted by one site.
We have  chosen the overlap real and positive for convenience.

%
%

As Fig.\ \ref{fig:size} shows, the overlaps
in the ISE and Heisenberg models are not only finite in the
$N\rightarrow \infty$ limit, but also surprisingly close to each other, despite
the fact that the ranges of the interaction in both models are quite different.
We now present an analytical calculation in the ISE model that
gives some understanding of origin of  this finite number.
The groundstate wavefunction of the ISE model in a basis of local spins $\{
 |\sigma_1\cdots \sigma_M\rangle\}$, $\sigma_i = \pm \frac{1}{2}$ is
given by~\cite{Hald94}:
\begin{equation}
\Psi^{0}_{M}\{\sigma_1\cdots\sigma_M\} =
\prod_{i<j}^{M}(z_i-z_j)^{\delta_{\sigma_i ,\sigma_j}}
e^{\frac{\pi i}{2}\rm{sgn}(\sigma_i - \sigma_j)}.
\end{equation}
For the $N$ site ISE groundstate $M=N$ and $\{z_i\}\equiv C_N =
\{e^{\frac{2 \pi i
n}{N}} \}_{n=1}^{N}$, while for the $N-2$ site groundstate $M=N-2$ and $\{
z_i\} \equiv C_{N-2} =\{e^{\frac{2\pi i n}{N-2}} \}_{n=1}^{N-2}$
To compute the ISE
overlap we add $\sigma_{N-1},\; \sigma_{N}$, sitting in a singlet, to the
$N-2$ site groundstate.  This overlap is hard to
calculate since the $z_i$'s from both
sets are not commensurate with each other.  However, if we slightly deform the
set $C_{N-2}$ to be $\{e^{\frac{2\pi i n}{N}}\}_{n=1}^{N-2}$ and leave
$\sigma_{N-1},\sigma_{N}$ in a singlet then we obtain a new state $\Psi^2_N$,
that can be recognized as a {\em localized}
2-{\em spinon state}~\cite{Hald91,TH94}.
Spinons are the elementary excitations of the ISE model with semionic
statistics~\cite{Hald91}.  Although this localized spinon state is not an
energy eigenstate it consists of an admixture of eigenstates that contain only
0 spinons (the groundstate on $N$ sites) or 2 spinons.  Here the localized {\em
spinons} sit at sites $N-1$ and $N$ in a singlet.  In general we could have
put them at sites $\alpha,\beta$ by deforming $C_{N-2}$ into $\{e^{\frac{2\pi i
n}{N}}|i=1,\ldots,N\}/\{e^{\frac{2\pi i\alpha}{N}},e^{\frac{2\pi
i\beta}{N}}\}$.

In the basis of states with
$M$ overturned spins with respect to the ferromagnetic state
labeled by their positions along the chain: $n_1,\ldots,n_M$,
$\Psi^0_N$ and $\Psi^2_{N}$
are given by:
\begin{eqnarray}
\Psi^0_N\left(n_1,\ldots,n_{\frac{N}{2}}\right) &=& \prod_{i=1}^{\frac{N}{2}}
(-)^{n_i}
\prod_{i<j}^{\frac{N}{2}} \sin^2\left(\frac{n_i -n_j}{N}\pi\right)\\
\Psi^2_N\left(n_1,\ldots,n_{\frac{N}{2}-1}\right)
&=&
\prod_{i=1}^{\frac{N}{2}-1} (-)^{n_i}
\prod_{i<j}\sin^2\left(\frac{n_i-n_j}{N}\pi\right) \nonumber\\
&&
\hspace{-.5truein}\times \prod_{i=1}^{\frac{N}{2}-1}
\sin\left(\frac{n_i-\alpha}{N}\pi\right)
\sin\left(\frac{n_i-\beta}{N}\pi\right)
\end{eqnarray}

We now calculate the overlap between $\Psi^0_N$ and $\Psi^2_N$ for arbitrary
separation $\alpha$ between the spinons (because of translational invariance
we can fix $\beta$ to be
0). We have for the overlap:
\begin{equation}
\frac{\langle\Psi^0 | \Psi^2(\alpha)\rangle}{\sqrt{
\langle\Psi^0 |\Psi^0\rangle \langle\Psi^2(\alpha) |\Psi^2(\alpha)\rangle}} =
\frac{\langle\Psi^0 |\Psi^2(\alpha)\rangle}{
\langle\Psi^0 |\Psi^0\rangle}
\left(\frac{\langle\Psi^2(\alpha) |\Psi^2(\alpha)\rangle}{\langle\Psi^0
|\Psi^0\rangle}\right)^{-\frac{1}{2}}
\end{equation}
First we will determine $\langle\Psi^0|\Psi^2(\alpha)\rangle$. Setting
$M=\frac{N}{2}-1$:
\begin{eqnarray}
 \langle\Psi^0|\Psi^2(\alpha)\rangle & = &
\left( \frac{N}{2} \right) !
 ~\sum_{ n_1,\ldots,n_M }
\Psi^2(n_1,\ldots,n_M|0,\alpha)\times\nonumber\\
&&\left\{\Psi^0(n_1,\ldots,n_M,0) -
\Psi^0(n_1,\ldots,n_M,\alpha)\right\}^* \nonumber\\
&=&\left( \frac{N}{2} \right) !~
\sigma(\alpha)\left(1-(-)^\alpha\right).
\label{sigmadef}
\end{eqnarray}
Here
\begin{eqnarray}
\sigma(\alpha) &=&\sum_{n_1,\ldots,n_M}
\left(\prod_{i<j}\sin^4\left(\frac{n_i-n_j}{N}\pi\right)\right)\nonumber\\
&\times&\prod_{i=1}^{M}\sin^3\left(\frac{\pi n_i}{N}\right)
\prod_{i=1}^{M}\sin\left(\frac{n_i-\alpha}{N}\pi\right).
\end{eqnarray}
In eq.\ (\ref{sigmadef}) we shifted all the $n_i$ by $\alpha$ in the second
term to bring both terms in the same form.

Since $\Psi^0$ is a singlet we know that if the two spinons in $\Psi^2$
were in the $S_z=0$ {\em triplet} state the overlap should be zero, i.e.\
$\langle\Psi^0|\Psi^2(\alpha),{\rm triplet}\rangle \propto
\sigma(\alpha)\left(1+(-)^\alpha\right) = 0$ for all $\alpha$.
Therefore $\sigma(\alpha)$ vanishes for all {\em even} $\alpha$. At the same
time we see from eq.\ (\ref{sigmadef}) that $\sigma(\alpha)$ is a polynomial
in $\cos\left(\frac{\pi\alpha}{N}\right)$. This is easily checked by
expanding the $\sin\left(\frac{n_i-\alpha}{N}\pi\right)$ and noting that
$\sigma(\alpha)$ is even in $\alpha$. We conclude
immediately:
\begin{eqnarray}
\sigma(\alpha)&\propto &\prod_{j=1}^{\frac{N}{2}}
\left(\cos\left(\frac{\pi\alpha}{N}\right)-\cos\left(\frac{2\pi
j}{N}\right)\right)\nonumber\\
&\propto & \frac{\sin\left(\frac{\pi\alpha}{2}\right)}{
\sin\left(\frac{\pi\alpha}{N}\right)}.
\end{eqnarray}
Now we turn our attention to the second piece.
Only its numerator is $\alpha$ dependent:
\begin{eqnarray}
\lefteqn{ \langle \Psi^2(\alpha)|\Psi^2(\alpha)\rangle =}\nonumber\\
&&2 \left(\frac{N}{2} -1\right)! ~\sum_{\{ n_1,\ldots,n_M\} }
\left(\Psi^2(n_1,\ldots,n_M|0,\alpha)\right)^2\nonumber\\
&&= 2\left(\frac{N}{2} -1\right)! ~\sum_{n_1,\ldots,n_M}
\prod_{i<j}\sin^4\left(\frac{n_i-n_j}{N}\pi\right)\times\nonumber\\
&&\hspace{.15truein}\prod_{i=1}^M\sin^2\left(\frac{\pi
n_i}{N}\right)\sin^2\left(\frac{n_i-\alpha}{N}\pi\right).
\end{eqnarray}
But this sum can be recognized as the $\langle S_z(\alpha) S_z(0)\rangle$
static correlation function in the ISE
model~\cite{Hald88}, or---after an (exact)
conversion of the sums to integrals---as the one-particle density matrix in the
Calogero-Sutherland model at half filling~\cite{Suth71}.  The result for large
$N$ is that
%
%
is proportional to
$\frac{{\rm Si}(\pi\alpha)}{\pi\alpha}$, where
${\rm Si}(x)$ is the sine-integral function.
The entire expression for the overlap, with normalization
computed by considering $\alpha =0$, then becomes:
\begin{equation}
\frac{2}{N}\frac{\sin\left(\frac{\pi\alpha}{2}\right)}{\frac{\pi\alpha}{N}}
\sqrt{\frac{\pi\alpha}{{\rm Si}(\pi\alpha)}}.
\label{twospinonov}
\end{equation}
Thus for a nearest neighbor the overlap is $\frac{2}{\pi}\sqrt{\frac{\pi}{
{\rm Si}(\pi)}}\simeq 0.82917$, which is within 1.5\% of the Heisenberg and ISE
groundstate singlet insertion overlaps.

Many of the properties of the two spinon matrix element can
be understood in terms of the Girvin-MacDonald-Read \cite{FQHE}
order parameter for the bosonic $\nu = \frac{1}{2}$
Laughlin \cite{Laugh} state.
In the spinon language,
the insertion of up-spin spinons acts on
the down-spin wavefunction in that same way as the quasi-hole
operator of the $\nu = \frac{1}{2}$
state, while the insertion of a down-spin spinon acts analogously
to the insertion of a quasi-hole and a hard-core boson.
Consequently, the two spinon insertion for
$\alpha = 0$ clearly takes the bosonic $\nu = \frac{1}{2}$ state
with $\frac{N}{2}-1$ particles,
periodic on a ring of size $N-2$, to the
$\nu = \frac{1}{2}$
state with $\frac{N}{2}$ particles
periodic on a ring of size $N$, and
is essentially the Girvin-MacDonald-Read
order parameter for the $\nu = \frac{1}{2}$
state and the overlap associated with it must
be identically
equal to one. Further, for $\alpha \neq 0$,
the two spinon insertion is equivalent to
a sum of two operators: a
%
quasi-hole
being inserted a finite distance from a hard core
boson and another quasi-hole.
 Such an object should still have an
expectation value given essentially by the
quasihole propagator. This would decay exponentially
in the bulk of a
$\nu = \frac{1}{2}$
state, but near the edge would decay only as
$\alpha^{-\frac{1}{2}}$, exactly as the two
spinon matrix element does (see Eq.\ (\ref{twospinonov})
and Fig.\ \ref{fig:sep}).
The vanishing of
the matrix element for any even $\alpha$  and
its alternation for odd $\alpha$ simply reflect
the lattice structure of our model,
the phase of the quasi-hole propagator
and the fact that the spinon operator is a
superposition of two operators so that the
result is real, rather than complex. The
reason why the two spinon overlap agrees so well with
the actual singlet insertion overlap is unclear, however,
our numerical Monte Carlo results for the singlet insertion display
the same power law decay as a function of the
spin separation (see figure~\ref{fig:sep}).
The insertion overlap also vanishes for separations which are
even multiples of the lattice spacing and alternation
for separations which are
odd multiples of the lattice spacing, so it would appear
that for the Heisenberg and ISE models the
spin chain ODLRO is essentially that same as the
ODLRO of the $\nu = \frac{1}{2}$  Laughlin state.

Note that, since
the two spinon insertion operator generates states with either
zero or two spinons, the zero spinon state being the
ground state,  its action is analogous to
the action of a pair of fermionic creation and annihilation
operators on the fermionic groundstate, generating
 states which contain one particle and one hole or
else no particles and holes, i.e. the groundstate.
The matrix element to the ground state as a function of
the space-time separation of the fermionic creation and
annihilation operators is the amplitude for the
fermion to propagate to the space-time location
where it can be annihilated and is by definition the
fermion propagator.
The overlap we calculate
is, by analogy, proportional to the equal time spinon propagator,
i.e.\ the amplitude for either of
the spinons to propagate to the location of the other,
at which point the two can annihilate each other, provided
they are in a singlet configuration.

In an effort to better understand our results
for the Heisenberg and ISE models, we
examined the overlap for the same insertion operator
acting on the $N-2$ site $XY$ model groundstate with the
$N$ site  $XY$ model groundstate.  The strikingly similar
result, here extended to a larger system size than
was possible for the Heisenberg and ISE models,  is
shown in Fig. \ref{fig:size}.  Since the $XY$ model can
be mapped onto spinless fermions, the ``order parameter''
can be recast in that language and a more detailed study made.
In that language the non-zero overlap is between the
 groundstate
for $N/2$ spinless fermions,
with
either periodic or
%
%
antiperiodic boundary conditions,
on $N$ sites
(with positive
hopping integral so that their momenta
%
%
in the groundstate are centered around
$\pi$)
and the state obtained by adding two sites and
one fermion
(in a superposition of being on
the two added sites with a relative minus sign for the two
different sites)
 to the
groundstate for the $N/2-1$ spinless fermions on
$N-2$ sites---with the opposite boundary conditions
from the $N$ site case.
%
%
The change of boundary conditions is a
non-local operation in the fermion language, however
in the spin language it arises from a local operation, i.e.
the insertion of two additional spins in a singlet
configuration.

Since the wavefunctions for the two
fermion states to be
overlapped are those of free
particles, considerable progress can be made in
its computation.
In particular, it can be bounded rigorously from below.
The calculation is straightforward,
using simple properties of polynomials which roots lie
on the unit circle,
and the fact that the
corresponding momenta
 for the $N-2$ and $N$ site models differ only
by $O(1/N^2)$
near the Fermi surface.
We will
present the proof elsewhere.
Here we state only the result that the lower bound
obtained was $e^{-\frac{5}{2}} + O(N^{-1})$.

The non-zero overlap for the XY model is
particularly striking since we know that
the overlap between the $N$ site,
$\frac{N}{2}$ particle
groundstate and the state created by
inserting a single, localized electron into the
$N$ site,
$\frac{N}{2}-1$ particle
groundstate would vanish like $N^{-\frac{1}{2}}$,
while the overlap between the $N$ site,
$\frac{N}{2}$ particle
groundstates between models with
periodic and antiperiodic boundary conditions
vanishes like $N^{-1}$.

The vanishing of the
overlap for the insertion of a single, localized fermion
implies that no analog of the two spinon
calculation exists for the $XY$ model, since
the only way to make the momenta similar  in the
two overlapped states is
to make $\frac{N}{2} -1$ of them identical is which
case the overlap will vanish like $N^{-\frac{1}{2}}$.

It may appear that, since the ``order parameter''
we have defined
involves spin chains with different numbers of sites,
there will be no physical consequences to this form of
ODLRO and in fact there is, for example, no additional groundstate
degeneracy associated with the existence of this kind of order.
However,
order of the type we
find can have important consequences
for more general models than spin
chains. For example, we will now show that
the ODLRO of the Heisenberg model is responsible for the
leading contribution to the singlet pair susceptibility of the
one dimensional Hubbard model and is thus not without
potentially important physical consequences.

First, we make the connection between the ODLRO found for
the Heisenberg model and the equal time,
singlet pairing correlation function
of the
one dimensional Hubbard model in the limit as
$U \rightarrow \infty$.  In that limit the
groundstate wavefunction of the Hubbard model
is given by a product of spin and charge wavefunctions,
the latter being given by a spinless fermion determinant and
the former by a
Bethe Ansatz wavefunction for the
``squeezed'' Heisenberg model, i.e.
a Heisenberg model defined only on those
sites occupied by the spinless fermions
\cite{OgataShiba}.
For periodic boundary conditions and the
number of electrons equal to $4N +2$, $N$
an integer, the ground state wavefunction for
the Heisenberg model should be used.  The equal time
singlet pair correlation function is given
by the overlap of the groundstate with the
state obtained from the groundstate by removing
a nearest-neighbor, singlet pair of electrons at
sites which we can take to be $0$ and $1$ and then
inserting a earest-neighbor, singlet pair of electrons
at sites $j$ and $j+1$. Due to the hidden ODLRO
of the Heisenberg model, the spin wavefunction overlap
for a fixed charge-configuration has a  piece which
for large separation, $j$, is just given by a constant
times $(-1)^{n_j}$, where $n_j$ is the number of
spinless fermions found on sites between $1$ and
$j$ in that charge configuration.  This leads
to a contribution to the correlation function
which is given by the expectation value in the
spinless fermion groundstate of
$\Psi^{\dagger}(j+1)\Psi^{\dagger}(j)
\Psi(1) \Psi(0) (-1)^{\sum_{j>l>1}\Psi^{\dagger}(l)\Psi(l)}$,
similar to the correlation function for the alternating,
spin-spin correlation
function studied by Sorella, {\it et al.} \cite{Sor}.
The leading
asymptotic behavior of this
expectation value
 can be computed straightforwardly from
Abelian bosonization
and is given by
$A~\cos(k_Fx) x^{-\frac{5}{2}}$, where $k_F$ is the
$k_F$ of the spinless fermions
and $A$ is a cutoff dependent constant.  This agrees with
the predictions of the Luttinger liquid description
of the Hubbard model \cite{hubb_ll,Frahm},
provided that we take the
charge sector rediagonalization parameter,
$K_{\rho}$, to be $1/2$
and remember that the $k_F$ of the spinless fermions
is twice that of electrons with spin at the same filling
fraction.

In the Hubbard model, this contribution to the
singlet pair correlation functions arises because
the bosonized form of
$\psi_{\uparrow}(j)\psi_{\downarrow}(j+1)$ contains an
operator proportional to
$\exp \left( i  \Theta_{R,\rho} \right)$, involving only charge
degrees of freedom.~\cite{notation}
This operator is present because
the operator product expansion for
$\exp \left( \frac{i}{2} \Theta_{R,\sigma}(x)\right)
\exp \left( -\frac{i}{2} \Theta_{R,\sigma}(x^{\prime})\right)$
contains the identity times a coefficient asymptotically
proportional to
$(x-x^{\prime})^{-\frac{1}{2}}$.
The decay of this coefficient with separation
implies that, if the electrons are inserted
$m$ sites apart where $m \gg 1$ but $j \gg m$,
the singlet pair correlations decay with $j$  in the same way,
 but
there is a multiplication of
the prefactor, $A$, by  a factor of $m^{-\frac{1}{2}}$
arising from spin degrees of freedom \cite{careful}.
This is in in agreement
with the decay we find for
the singlet and two spinon insertions in the ISE and
Heisenberg models, and suggests the identification
of the
insertion of an up-spin spinon into the ISE model
with the action of the operator
$O_{insert} =
i^j~\exp \left( \frac{i}{2} \Theta_{R,\sigma}(j~a)\right)
+(-i)^j~\exp \left( \frac{i}{2} \Theta_{L,\sigma}(j~a)\right)$,
where $a$ is the lattice spacing and $j$ is the number of sites
to the left of the insertion site {\it in the original chain}.
Since this operator is a semion, it is natural to identify
it as the spinon creation operator; an identification
compatible
with the observation of \cite{ludwig} that the generalized
commutation relations of the
Fourier modes
of this operator provide a natural realization of the Yangian.

The identification is also compatible with the alternation
with odd separation and vanishing for
even separation that we find for the singlet and two
spinon insertions.
Both are in agreement with the singlet pair correlations of the Hubbard model
as computed with Abelian bosonization.

The connection to Abelian bosonization should
be extendable to generalizations of the Hubbard model
which include
spin-dependent interactions so that
the Luttinger liquid
rediagonalization parameter of the
 spin sector of the Hubbard model, $\kappa_{\sigma}$,
(for a definition of $\kappa_{\sigma}$ see \cite{hubb_ll})
is renormalized from $1$.
At half filling
the low energy sector of
the  model then becomes a general $XXZ$ model.
This would change the exponent
for the decay of the singlet insertion
overlap from $\frac{1}{2}$ to
an exponent,
$\eta = \frac{1}{4}(\kappa_{\sigma} + \kappa_{\sigma}^{-1})$.
For $XXZ$ models
this is a simple function of the anisotropy of the
model \cite{Affleck}.
In general: $\kappa_\sigma = 1-\frac{1}{\pi}\cos^{-1}\left(\frac{J_z}{J_{xy}}
\right)$.
This is equal to
$\frac{5}{8}$ at the
$XY$ point exactly as we observe.
Based on the connection between
the two spinon overlap  and
the singlet insertion for the ISE model, where spinons are
noninteracting, we argued that
the decay exponent was just
the exponent of the spinon-spinon propagator. In this language,
the continuously varying exponent of the overlap decay
in general $XXZ$ chains
is just the exponent of the spinon-spinon propagator in
models of interacting spinons.

We note that
that the alternating piece in the singlet-pair
correlation function is the slowest decaying
piece of that
correlation function for the Hubbard model
and may have important consequences
for superconducting correlations in models based on the
one dimensional Hubbard model.
In particular, if one considers an array of Hubbard chains
coupled with operators
which properly correlate the positions of the electrons  on
neighboring
chains without disrupting the singlet
overlap property
or the charge degrees of freedom greatly then the scaling
dimension of pair hopping between the chains will be
renormalized and pair hopping will become relevant and
lead to an instability similar to that
envisioned in the interlayer tunneling
mechanism for superconductivity.
Relevant operators having this effect occur naturally
when the spin-spin superexchange interaction
between chains is considered in the bosonization language
\cite{us}.


We gratefully acknowledge
helpful correspondences with
M.~Ogata and discussions with
F.~D.~M.~Haldane and D.~G.~Clarke,
as well as financial support from
NSF grants DMR-9104873 (P.~W.~A.) and
DMR-922407 (J.~C.~T.) and
the NEC corporation (S.~P.~S.).

\nopagebreak


%
%
\begin{figure}
\caption{ Calculated Overlap for the ISE and Heisenberg Models}
Shown are the calculated overlaps between the $N$ spin groundstate of
the $1/r^2$, nearest neighbor Heisenberg and $XY$ models
and the states obtained by inserting a nearest
neighbor singlet pair of spins into the $N-2$ spin groundstates of those
models.  The dashed line is
a fit to the ISE results using
$0.0817 + 0.778 N^{-2}$, the solid to Heisenberg results
using $0.820 + 0.740 N^{-2}$,
and the dotted and dashed
a fit to the $XY$ results using $ 0.808 + .819 N^{-2}$.
%
%
%
%
\label{fig:size}
\end{figure}

 \begin{figure}
 \caption{ The Decay with Separation of the Overlap}
The exact overlap for the $XY$ and the Monte Carlo
calculated overlap for the $ISE$ model
as  functions of the system size
for spins inserted in a singlet configuration separated by
half the size of the system. The results obtained in this way
should be purely power law in the large system size limit,
whereas correlation functions at fixed system size generally
deviate from power law when the separation is not small compared to
the system size.  We have also examined correlation functions
at fixed system size,
and exact results for smaller systems
for the ISE and Heisenberg models. All of those
produced results consistent with
$\alpha^{-\frac{1}{2}}$ behavior  for the ISE and Heisenberg models
and $\alpha^{-\frac{5}{8}}$  behavior  for the $XY$ model.
 \label{fig:sep}
 \end{figure}

%
%


\begin{references}

\bibitem{pwa_conj} P.~W.~Anderson, {\it Princeton RVB Book},
unpublished.

\bibitem{Hald88} F.D.M.\ Haldane, Phys.\ Rev.\ Lett.\ {\bf 60}, 635 (1988),
B.S.\ Shastry, Phys.\ Rev.\ Lett.\ {\bf 60}, 639 (1988).

\bibitem{FQHE} S.~Girvin and A.~MacDonald, Phys. Rev. Lett. {\bf 58},
1252 (1987); N.~Read, Phys. Rev. Lett. {\bf 62}, 86 (1989).

\bibitem{PS} Thomas Pruschke and Hiroyuki Shiba, Phys. Rev. {\bf B46},
356 (1992).

\bibitem{Hald94} F.D.M.\ Haldane in {\em Proc.  of the 16th
Tanaguchi Symposium on Condensed Matter, Kashikojima, Japan, Oct.  26-29,
1993},
edited by A.  Okiji
and N.  Kawakami (Springer-Verlag, Berlin-Heidelberg-New York, 1994), {\bf
cond-mat 9401001}.

\bibitem{Hald91} F.D.M.\ Haldane, Phys.\ Rev.\ Lett.\ {\bf 66}, 1529
(1991).

\bibitem{TH94} J.C.\ Talstra and F.D.M.\ Haldane, Phys.\ Rev.\ B {\bf 50}
6889 (1994).



\bibitem{Suth71} B.\ Sutherland, Phys.\ Rev.\ A {\bf 4}, 2019 (1971).

\bibitem{Laugh} R.~B.~Laughlin, Phys. Rev. Lett. {\bf 50}, 1395 (1983).

\bibitem{Haldane} F.~D.~M.~Haldane, Phys. Rev. Lett. {\bf 47}, 1840 (1981);
F.~D.~M.~Haldane, J. Phys. C {\bf 14}, 2585 (1981).

\bibitem{OgataShiba} M.~Ogata and H.~Shiba, Phys. Rev. {\bf B41}, 2326
(1990).

\bibitem{Sor} A.~Parolla and S.~Sorella, Phys. Rev. Lett. {\bf 64}, 1839
(1990).

\bibitem{notation}
Here $\exp \left( i  \Theta_{R,\uparrow} \right)$ creates a right moving,
up spin
fermion and $\Theta_{R,\rho}$ and $\Theta_{R,\sigma}$ are the
symmetric and antisymmetric combinations of the up and down spin
phase fields.

\bibitem{careful} This is more than offset by a charge
degree of freedom induced renormalization
of $A$ by $m^{+\frac{5}{8}}$, which can be derived from
either the Ogata-Shiba wavefunction
or the Luttinger liquid description of the large $U$
Hubbard model.


\bibitem{ludwig} P.~Bouwknegt, A.~W.~W.~Ludwig, K.~Schoutens,
Phys.\ Lett.\ B, in press, and hep-th {\bf 9406020}.

\bibitem{hubb_ll} H.~J.~Schulz, Phys. Rev. Lett. {\bf 64}, 2831
(1989); S.~Sorella, A.~Parolla, M.~Parrinello and E.~Tosatti,
Europhys. Lett. {\bf 12}, 729 (1990); H.~Shiba and M.~Ogata,
Prog. Theor. Phys. Supplement {\bf 108}, 265 (1992).

\bibitem{Frahm} Holger Frahm and V.~E.~Korepin, Phys. Rev. {\bf B42},
10533 1990.

\bibitem{Affleck}  I. Affleck, {\it Fields, Strings and Critical
Phenomena-Les Houches, Session XLIX (1988)}, edited by E. Br\'{e}zin
and J. Zinn-Justin (Elesvier Science Publishers B. V., 1989).

\bibitem{us} S.~P.~Strong, J.~C.~Talstra and P.~W.~Anderson, unpublished.


\end{references}
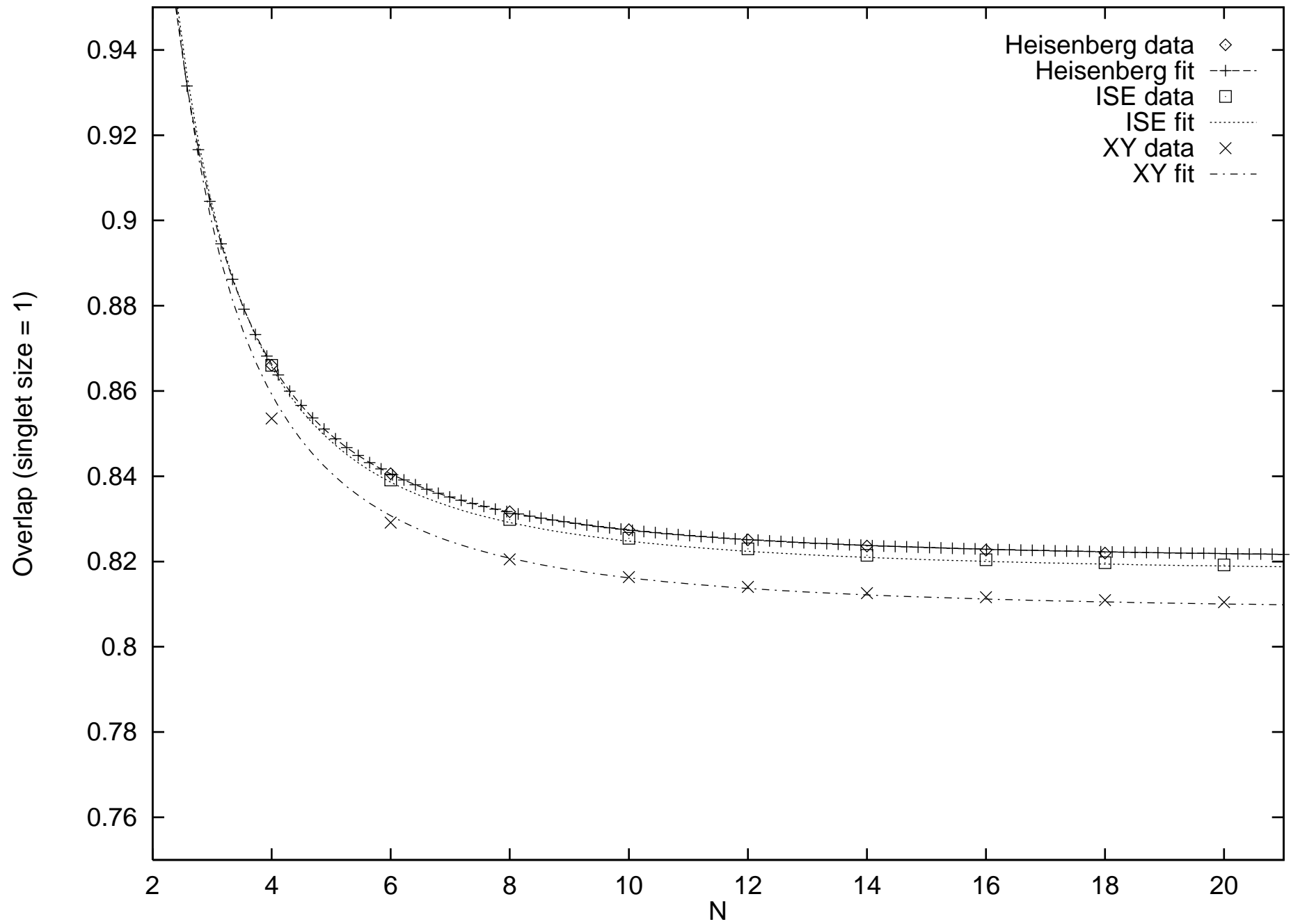
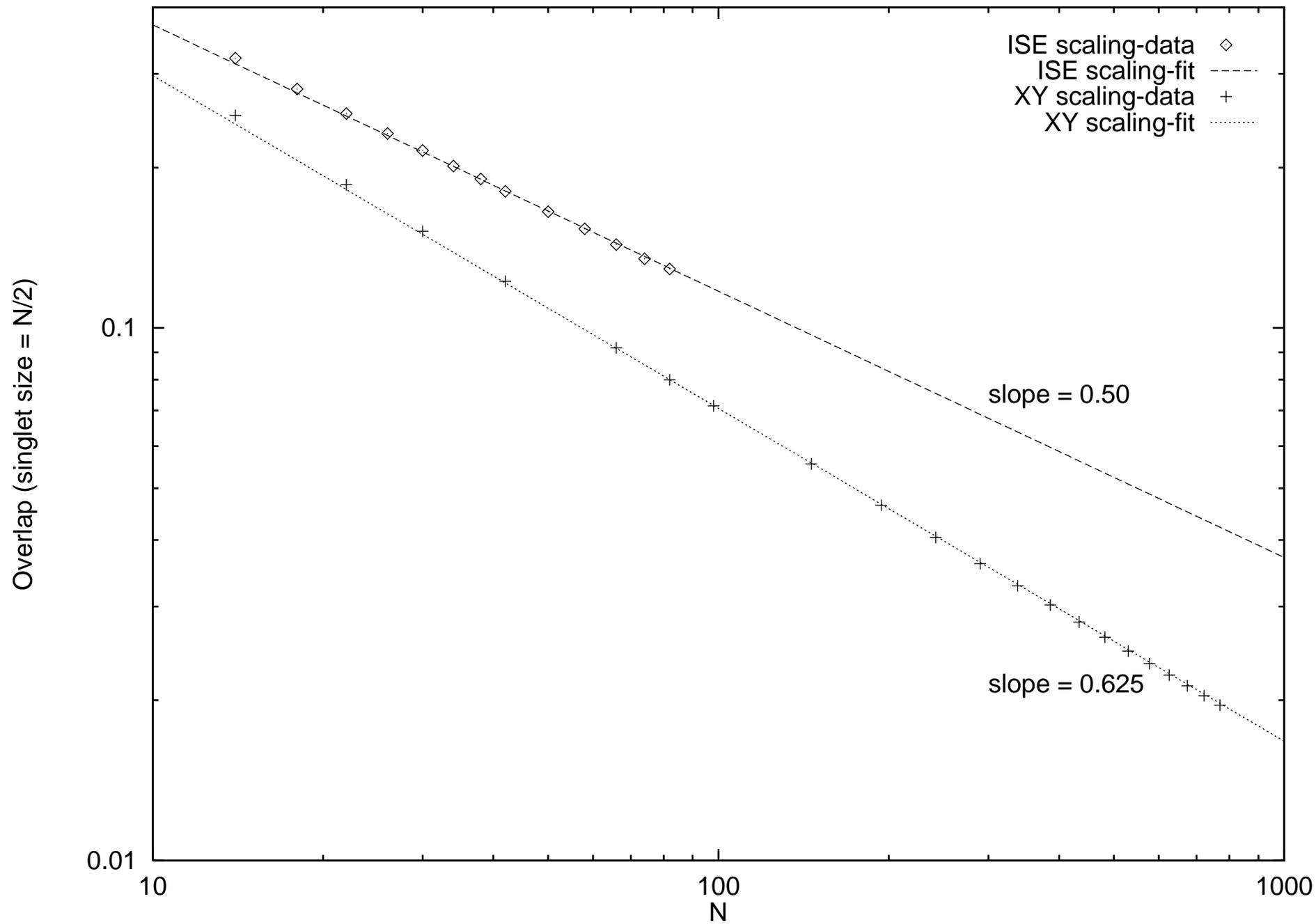
\end{document}